\documentclass[prl,twocolumn,letterpaper,superscriptaddress,footinbib]{revtex4}
\include{epsf}
\usepackage{graphicx}
\usepackage{pslatex}

\begin{document}

\title{Constraints on Cosmic Neutrino Fluxes from the ANITA Experiment}

\author{S.~W.~Barwick}\affiliation{Department of Physics and Astronomy, University of California at Irvine, Irvine, California}
\author{J.~J.~Beatty}\affiliation{Department of Physics, Ohio State University, Columbus, Ohio}
\author{D.~Z.~Besson}\affiliation{Department of Physics and Astronomy, University of Kansas, Lawrence, Kansas}
\author{W.~R.~Binns}\affiliation{Department of Physics, Washington University in St. Louis, St. Louis, Missouri}
\author{B.~Cai}\affiliation{School of Physics and Astronomy, University of Minnesota, Minneapolis, Minnesota}
\author{J.~M.~Clem}\affiliation{Bartol Research Institute, University of Delaware, Newark, Delaware}
\author{A.~Connolly}\affiliation{Department of Physics and Astronomy, University of California at Los Angeles, Los Angeles, California}
\author{D.~F.~Cowen}\affiliation{Department of Astronomy and Astrophysics, Pennsylvania State University, University Park, Pennsylvania}
\author{P.~F.~Dowkontt}\affiliation{Department of Physics, Washington University in St. Louis, St. Louis, Missouri}
\author{M.~A.~DuVernois}\affiliation{School of Physics and Astronomy, University of Minnesota, Minneapolis, Minnesota}
\author{P.~A.~Evenson}\affiliation{Bartol Research Institute, University of Delaware, Newark, Delaware}
\author{D.~Goldstein}\affiliation{Department of Physics and Astronomy, University of California at Irvine, Irvine, California}
\author{P.~W.~Gorham}\affiliation{Department of Physics and Astronomy, University of Hawaii at Manoa, Honolulu, Hawaii}
\author{C.~L.~Hebert}\affiliation{Department of Physics and Astronomy, University of Hawaii at Manoa, Honolulu, Hawaii}
\author{M.~H.~Israel}\affiliation{Department of Physics, Washington University in St. Louis, St. Louis, Missouri}
\author{J.~G.~Learned}\affiliation{Department of Physics and Astronomy, University of Hawaii at Manoa, Honolulu, Hawaii}
\author{K.~M.~Liewer}\affiliation{Jet Propulsion Laboratory, Pasadena, California}
\author{J.~T.~Link}\affiliation{Department of Physics and Astronomy, University of Hawaii at Manoa, Honolulu, Hawaii}
\author{S.~Matsuno}\affiliation{Department of Physics and Astronomy, University of Hawaii at Manoa, Honolulu, Hawaii}
\author{P.~Mio\v{c}inovi\'c}\affiliation{Department of Physics and Astronomy, University of Hawaii at Manoa, Honolulu, Hawaii}
\author{J.~Nam}\affiliation{Department of Physics and Astronomy, University of California at Irvine, Irvine, California}
\author{C.~J.~Naudet}\affiliation{Jet Propulsion Laboratory, Pasadena, California}
\author{R.~Nichol}\affiliation{Department of Physics, Ohio State University, Columbus, Ohio}
\author{K.~Palladino}\affiliation{Department of Physics, Ohio State University, Columbus, Ohio}
\author{M.~Rosen}\affiliation{Department of Physics and Astronomy, University of Hawaii at Manoa, Honolulu, Hawaii}
\author{D.~Saltzberg}\affiliation{Department of Physics and Astronomy, University of California at Los Angeles, Los Angeles, California}
\author{D.~Seckel}\affiliation{Bartol Research Institute, University of Delaware, Newark, Delaware}
\author{A.~Silvestri}\affiliation{Department of Physics and Astronomy, University of California at Irvine, Irvine, California}
\author{B.~T.~Stokes}\affiliation{Department of Physics and Astronomy, University of Hawaii at Manoa, Honolulu, Hawaii}
\author{G.~S.~Varner}\affiliation{Department of Physics and Astronomy, University of Hawaii at Manoa, Honolulu, Hawaii}
\author{F.~Wu}\affiliation{Department of Physics and Astronomy, University of California at Irvine, Irvine, California}

\begin{abstract}
We report new limits on cosmic neutrino fluxes
from the test flight of the Antarctic Impulsive Transient Antenna
(ANITA) experiment, which completed an 18.4 day flight of a prototype 
long-duration balloon payload, called {\em ANITA-lite}, in early 2004. 
We search for impulsive events that could be associated with 
ultra-high energy neutrino interactions
in the ice, and derive limits that constrain several
models for ultra-high energy neutrino fluxes and rule out the long-standing
Z-burst model. We set a 90\% CL integral flux limit on a pure $E^{-2}$
spectrum for the energy range $10^{18.5}$~eV$\leq E_{\nu} \leq 10^{23.5}$~eV at
$E_{\nu}^2 F \leq 1.6 \times 10^{-6}$ GeV cm$^{-2}$ s$^{-1}$ sr$^{-1}$.
\end{abstract}

\maketitle

Cosmic rays of energy above $3 \times 10^{19}$~eV are almost certain to be 
of extragalactic origin. Their gyroradius far exceeds that required
for magnetic confinement in our galaxy. At this energy
pion photo-production losses on the cosmic microwave background radiation (CMBR)
via the Greisen-Zatsepin-Kuzmin (GZK~\cite{GZK}) process 
limit their propagation distances to the local supercluster, 
of order 40~Mpc or less. However, the neutrinos that result from
this process~\cite{BZ70} would be observable out to the edge of
the visible universe. Recent studies 
make compelling arguments that input from neutrino observations
will be necessary to resolve the ultra-high energy
cosmic ray (UHECR) problem~\cite{Seck05}.
Neutrinos are coupled to the highest energy cosmic rays both
as a direct byproduct, and perhaps as a potential source of them.
Straightforward reasoning indicates there is a required cosmogenic 
neutrino flux~\cite{BZ70} with a broad peak in the energy range of $10^{17-19}$~eV.
First, Lorentz invariance allows transformation  of the cross section 
for photo-pion production at center-of-mass energies of order 1~GeV,
the $\Delta^+$-resonance energy, up to GZK energies, a boost of order $10^{11}$.
Second, precision measurements of the CMBR establish its flux density
for all cosmic epochs and thus determine the number density of boosted targets
for the photo-pion production process. Third, we apply the 
standard cosmological postulate that the cosmic-ray sources
are not uniquely overdense (and hidden) in our local supercluster
compared to the cosmic distribution.
Finally, evidence from composition studies indicates that the
UHECRs are hadronic, and thus unable to
evade interaction with the CMBR, 
even if they are as heavy as iron~\cite{Abb05}.
We conclude that 
any localized source of UHECR
at any epoch is surrounded by a ``GZK horizon'' beyond which the opacity of the CMBR to 
photo-pion interactions is sufficient to completely attenuate the charged 
progenitors, yielding pion secondaries which decay to neutrinos.
The intensity of all of these GZK neutrino spheres sums
to a quasi-isotropic cosmogenic neutrino flux, unless any of the assumptions
above is strongly violated, which
would in turn constitute a serious departure from Standard Model physics. 

Neutrinos may not only be cosmogenic byproducts, but could also 
be closely associated with sources of the UHECR, 
though this possibility is far more speculative.
If there are large fluxes of neutrinos at energies of order $10^{22-23}$~eV,
they can annihilate with Big-Bang relic cosmic background 
neutrinos ($T_{\nu}\sim 1.9$K) in our own Galactic halo via the interaction
$\nu \bar{\nu} \rightarrow Z^0$, the {\em Z-burst} process~\cite{Fod02,Kal02a,Wei99,Wei82}.
Decays of the neutral weak vector boson $Z^0$ then yield UHECRs,
overcoming the GZK cutoff because of the nearby production.
Moreover, Topological Defect (TD) models~\cite{Yos97} 
postulate a flux of super-heavy ($10^{24}$~eV) 
relic particles that decay in our current epoch and within the Earth's
GZK sphere, yielding both neutrinos and UHECR hadrons in the process.
Variant versions of such models, including hypothetical 
mirror-matter~\cite{MirrorNeck},  
can evade standard bounds to TD models; such variants
currently have the weakest experimental constraints.
Limits of the fluxes of ultra-high energy (UHE) neutrinos can 
constrain or eliminate such models as sources for the
UHECR. Both of these classes of neutrino models predict fluxes
well above the current predictions for cosmogenic GZK neutrinos.
In all models, the neutrino fluxes
in the $10^{18-20}$~eV energy range are well below what can be observed
with a cubic kilometer target volume, so detection methods must
use larger scales.

The ANITA mission is
now completing construction for a first launch as a 
long duration balloon payload in 2006. The mission
has a primary design goal of detecting EeV cosmogenic neutrinos, 
or providing a compelling flux limit.
ANITA will detect neutrino interactions through coherent radio Cherenkov
emission from neutrino-induced electromagnetic (EM) particle 
cascades within the ice sheet. 
The ANITA-lite prototype flew as a piggyback instrument aboard the
Trans-Iron Galactic Element Recorder (TIGER) 
payload. The payload launched Dec. 18, 2003, and was aloft
for 18.4 days, spending a net 10 days over the ice in its 1.3 circuits
of Antarctica. The payload
landed on the ice sheet several hundred km from
Mawson Station (Australia) at an elevation of 2500~m. 
ANITA-lite investigated possible backgrounds to
neutrino detection in Antarctica and verified many of the
subsystems to be used by the full-scale ANITA.
The payload operation was successful, and we have searched
for neutrino-induced cascades among the impulsive events measured.
While ANITA-lite did not have adequate directional resolution
to establish with certainty that the origin of any event was
within the ice, the data quality was sufficient to 
distinguish events that were consistent with 
cascades, and exclude events which were not, thus enabling
us to establish flux limits in the absence of candidate events.

ANITA exploits a property of EM cascades that has become known as
the Askaryan effect~\cite{Ask62}. During the development of
the EM cascade, selective electron scattering processes lead
to a negative charge asymmetry, inducing
strong coherent radio Cherenkov radiation in the form
of impulses with unique broadband spectral and polarization properties.
When a high energy neutrino showers in the ice sheet,
which has a radio attenuation length $L_{\alpha}\geq 1$~km~\cite{icepaper}, 
the resulting impulses can easily propagate up through the surface
to the balloon payload.  From balloon
altitudes of 37~km, the horizon is at nearly 700~km
distance, giving a synoptic view of  $\sim 1.5$~M km$^2$
of ice, or $\sim 2$~M km$^3$ volume to a depth of $\simeq L_{\alpha}$.
ANITA will consist of a $2\pi$ array of dual-
polarization antennas designed to monitor this entire
ice target. ANITA-lite flew only two first-generation ANITA antennas,
with a field-of-view covering about 12\% of the $ 1.5$~M km$^2$ ice sheet
area within its horizon at any time, but the $\approx 170,000$~km$^2$ 
area of ice in view still represents
an enormous monitored volume for the uppermost kilometer of ice
to which we were primarily sensitive. This leads to the strongest
current limit on neutrino fluxes within its energy regime.

The ANITA-lite antennas are dual-linear-polarization vertical (V) and
horizontal (H) quad-ridged horn antennas,
sensitive over 230-1200~MHz, over
which their angular response remains single-mode, with
a nearly constant effective directivity-gain of about 9-11 dBi. The
antenna beam is somewhat ellipsoidal in shape, with average
beamwidths of $59^{\circ}$ and $37^\circ$ in E-plane and H-plane, respectively.
The antenna boresights were offset from one another by 1~m lateral separation,
$22.5^{\circ}$ in azimuth, and were canted
$10^{\circ}$ downward in elevation.
The antennas were designed to retain an off-axis directivity of $\geq6$~dBi 
at the angles corresponding to the boresights of the
adjacent antennas; thus each antenna retains good sensitivity
to trigger on events that are centered on its nearest neighbor's
field-of-view. The combined field-of-view of the two antennas taken in
coincidence is of order $45^{\circ}$ in azimuth for typical
events, but can be significantly larger for strong impulses.

\begin{figure}
\begin{center}
\includegraphics[width=3.3in]{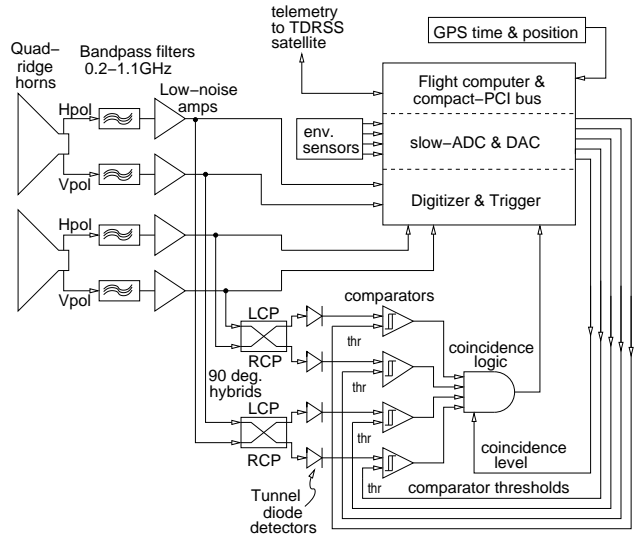}
\caption{The ANITA-lite system block diagram.}
\label{ALsystem}
\end{center}
\end{figure}

A block-diagram of the ANITA-lite antenna, trigger, and data acquisition 
system is shown in Fig.~\ref{ALsystem}. The H- and
V-polarization antenna voltages are first filtered to limit the passband to 0.2--1.1~GHz. A 
notch filter (not shown) removes a region around 400~MHz used by payload telemetry. The signals are
amplified by low-noise amplifiers (LNAs) with approximately 100~K noise figure,
for a net gain of $\sim 62$~dB.
The resulting signals, with thermal-noise levels corresponding
to $\sim 35$~mV rms, are
split equally between the digitizers and
the trigger coincidence section.

Coherent Cherenkov emission from showers in solid media is
100\% linearly polarized~\cite{Sal01}, and Antarctic ice does not
produce significant depolarization over the propagation
distances ($\sim 1$ km) required for 
detection of neutrino interactions~\cite{Birefringence}.
ANITA-lite takes advantage of this characteristic by requiring
that any trigger have roughly equal amplitude in left- and right-circular
polarizations (LCP \& RCP). This favors signals with
a high degree of linear polarization, and provides of order a factor of
two improvement in rejecting circularly-polarized
backgrounds. The conversion from
the H- and V-polarizations received from the antenna
into LCP and RCP is done by broadband $90^{\circ}$ hybrid-mode combiners.

The trigger system is critical to the 
sensitivity of a radio impulse detection system. It initiates
digitization of antenna waveforms based on correlated pulse amplitudes
among the different antenna channels. 
For ANITA-lite, the trigger required a 1- to 3-fold
coincidence among the four independent channels (two antennas
and two polarizations), where each channel was required to exceed
a power threshold during a 30~ns window.
The pulse-height spectrum
of received voltages due to ideal thermal noise is nearly Gaussian,
and ANITA-lite was operated with an average threshold
corresponding to $4.3~\sigma_V$, where 
$\sigma_V = \sqrt{k\langle T_{sys}\rangle Z\Delta\nu~}$ for
bandpass-averaged system temperature values of 
$\langle T_{sys}\rangle\approx 700$~K during
the flight. Here $k$ is Boltzmann's constant, $Z=50~\Omega$, and
$\Delta \nu = 800$~MHz is the system bandwidth.

\begin{figure}
\includegraphics[width=85mm]{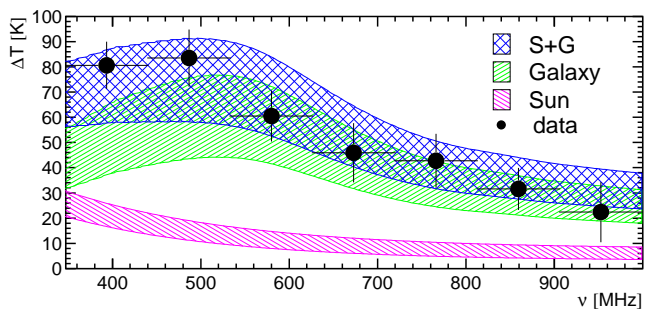}
\caption{\label{fig:ycal}(Color online) Frequency dependence of the 
excess effective antenna temperature $\Delta T$ when pointing to 
the Sun and the Galactic Center
~\cite{LowBandNoise}.
The top band is a model of the expected $\Delta T$, with a width 
equal to the systematic uncertainties. The lower two bands give 
contributions due to galactic and solar emissions, respectively. The antenna 
frequency response is folded into the model.}
\end{figure}

Calibration of the system gain, timing, and noise temperature was
performed by several means. 
A calibrated noise diode was coupled to the system between
the antenna and the first bandpass filter. Also, during the
first day of the flight, a pulse generator and transmitter antenna
at the launch site (Williams field, near McMurdo Station) 
illuminated the payload with
pulses synchronized to GPS signals. An onboard GPS signal
synchronously triggered the ANITA-lite system. These pulses
were recorded successfully by the system out to several hundred
km distance. Timing analysis of these signals indicates
that pulse phase could be estimated to a precision of 150~ps
for $\geq 4\sigma$ signal-to-noise ratio. Finally,
the response of the antennas to broadband noise from both
the Sun and the Galactic center and plane was determined by
differential measurements using data when the payload (which
rotated slowly during the flight) was toward or away from
a given source. The results of this analysis are shown in
Fig.~\ref{fig:ycal}, showing the spectral response function
with the various contributions from astronomical sources.
The ambient RF noise levels at balloon float altitudes were
found to be consistent with thermal noise due to the
ice at $T_{eff}\sim 250$~K and our receiver noise temperature
of $300-500$~K, which included contributions from the cables,
LNA, connectors, filters, and power limiters. 
Other than our own ground calibration 
signals, we also detected no sources of impulsive noise that
could be established to be external to our own payload.
Several types of triggers were investigated for correlations
to known Antarctic stations, and no such correlations were found.

\begin{figure}
\includegraphics[width=80mm]{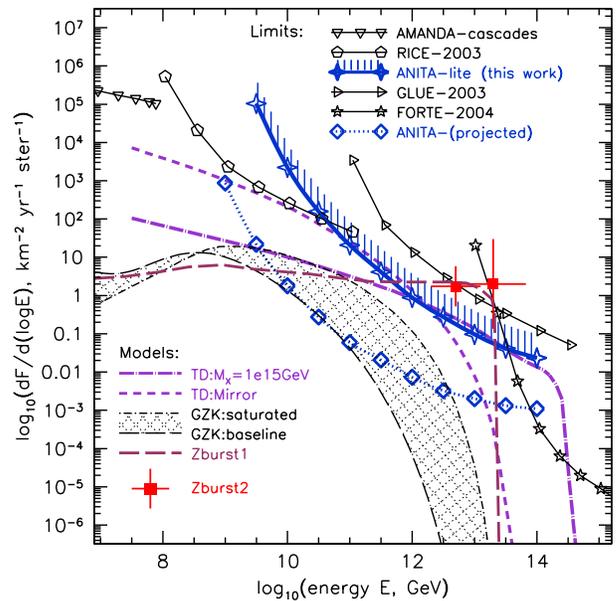}
\caption{\label{lim05}(Color online) Limits on various models
for neutrino fluxes at EeV to ZeV energies. The limits are: 
AMANDA cascades~\cite{AMcas},
RICE~\cite{RICE03},
the current work, GLUE~\cite{GLUE04}, 
the FORTE satellite~\cite{FORTE04},
and projected sensitivity for the full ANITA. Models
shown are Topological Defects for two values of the
X-particle mass~\cite{Yos97}, a TD model involving
mirror matter~\cite{MirrorNeck}, a range of models
for GZK cosmogenic neutrinos~\cite{Engel01,PJ96,Aramo05},
and several models for Z-bursts~\cite{Fod02,Kal02}. In 
the Z-burst models plotted as points, the flux is a
narrow spectral feature in energy, and the error-bars shown
indicate the range possible for the 
central energy and peak flux values. }
\end{figure}

ANITA-lite recorded $\sim 113,000$ events at an average livetime of
40\%~\cite{Livetime}. Of these events, $\sim 87,500$ are 3-fold-coincident
triggers considered for data analysis. The remainder were recorded for 
system calibration and performance verification. Two independent analyses 
were performed within collaboration, both searching for narrow Askaryan-like
impulses, in which almost all signal power is delivered within 10~ns about peak
voltage, time-coincident in at least two of four channels. Analysis A 
primarily relied on matched-filtering the data with the expected signal shape 
and requiring the filtered data to show better signal-to-noise ratio than 
the unfiltered data. 
Analysis B primarily relied on rejecting events which show high level of 
cross-correlation with known payload-induced noise events. 
This approach very efficiently removes the very common repetitive 
payload noise events. These constituted about 90\% of triggers, with
the remainder from unknown sources, probably also on the payload.
None of these resembled the expected neutrino signals. 
Analysis A determined the signal passing 
efficiency by tightening the cuts until the last background noise event was 
removed, and found 53\% of the simulated signal still passing the cuts.
Analysis B blinded 80\% of the data, optimized the cuts with the other 20\% 
using the model rejection factor technique~\cite{hill}, and found 65\% 
signal efficiency. No data events pass either of the analyses. In both 
analyses, the systematic uncertainty in passing rates was estimated at  
$\sim 20$\%. 

To estimate the effective neutrino aperture and exposure for
ANITA-lite, two different and relatively mature simulation codes
for the full ANITA instrument were modified to account for the
ANITA-lite configuration. These simulations account for propagation
of neutrinos through earth crust models, for the various interaction
types and neutrino flavors, for inelasticity, and both hadronic and
electromagnetic interactions (including LPM effects~\cite{Alv97}).
The shower radio emission
is estimated via standard parameterizations which have been validated
at accelerators~\cite{Sal01,SalSA}. Propagation through the ice uses a
frequency- and temperature-dependent model for $L_{\alpha}$, based on
data measured at the South Pole~\cite{icepaper}.
Surface refraction is accounted for using a combination of ray- and physical-optics. 
Refracted emission is propagated
geometrically to the payload where a detailed instrument model, based 
on lab measurements of the spare flight system, is applied to determine whether a 
detection occurs.

Based on the treatment described in Refs.~\cite{Anch02,FORTE04}, the
resulting model-independent 90\%~CL limit on neutrino fluxes with Standard 
Model cross-sections~\cite{Gan00} is shown in Fig.~\ref{lim05}.
ANITA-lite approaches the highest energy cosmogenic 
neutrino flux model~\cite{Aramo05}, and now appears to have entirely
excluded the Z-burst model~\cite{Eberle04,Fod02,Wei82} 
at a level required to account for the fluxes of the
highest energy cosmic rays, as represented by the two crosses in
the figure, with vertical and horizontal bars indicating  
the range of allowed model parameters for this case. Prior
limits from the GLUE and FORTE experiments had 
constrained most but not all of this range. 
Our limits rule out all of the remaining range for two of the highest
standard topological defect models, shown in Fig.~\ref{lim05}, 
both of which were constrained already by other experiments.
We also provide the first experimental limits on the highest mirror-matter 
TD model~\cite{MirrorNeck}. Table~\ref{table1} shows the 
expected event totals and limits for several of these models.
The ANITA-lite 90\% CL integral flux limit on a pure $E^{-2}$ spectrum
for the energy range $10^{18.5}$~eV $\leq E_{\nu} \leq 10^{23.5}$~eV is 
$E_{\nu}^2 F \leq 1.6 \times 10^{-6}$ GeV cm$^{-2}$ s$^{-1}$ sr$^{-1}$.

\begin{table}[htb!]
\caption{Expected numbers of events from few UHE neutrino models, and 
confidence level of exclusion by ANITA-lite observations.
\label{table1}}
\vspace{3mm}
 \begin{footnotesize}
  \begin{tabular}{|l|c|c|}
\hline
{ {\bf Model \& references}}   &  {\bf Events}      &  {\bf CL,\%} \\ \hline
{\it Baseline GZK models}~\cite{Engel01,PJ96,Aramo05} &0.009  & ... \\
\hline 
{\it Strong source evolution GZK models}~\cite{Engel01,Aramo05,Kal02} & 0.025-0.048 & ... \\
\hline
{\it GZK Models that saturate all bounds}~\cite{Kal02,Aramo05}: & 0.48 to 0.60 & 38 to 45\\
\hline
{\it {Topological Defects}:} & & \\
~~Yoshida {\it et al.} 1997, $M_X=10^{16}$~GeV~\cite{Yos97} &   7.8 &  99.959 \\ 
~~Yoshida {\it et al.} 1997, $M_X=10^{15}$~GeV~\cite{Yos97} &   22 &  100.000 \\ 
~~Berezinsky 2005, Mirror Necklaces~\cite{MirrorNeck} &   6.4 & 99.834 \\ 
\hline
{\it {Z-burst Models}:} & & \\
~~Fodor et al. (2002): Halo background~\cite{Fod02} & 5.0 &    99.326 \\
~~Fodor et al. (2002): Extragalactic background ~\cite{Fod02}& 14.2 &    99.999 \\
~~Kalashev et al. (2002)~\cite{Kal02a} & 45.9 & 100.000 \\ \hline
  \end{tabular}
 \end{footnotesize}
\end{table}

Although designed primarily as an engineering test, ANITA-lite has set 
the best current limits on neutrino fluxes above $10^{19.5}$~eV,
improving constraints by more
than an order of magnitude over the GLUE results~\cite{GLUE04}.
This demonstrates the power of the radio Cherenkov technique
applied to the balloon-based observations of the Antarctic
ice. Simulations for ANITA shown in Fig.~\ref{lim05}, 
indicate totals of order 5-50 events for the GZK model range
shown for 50 days of flight time, sufficient to detect
these model fluxes for the first time.

This work has been supported by NASA. We thank the Columbia Scientific
Balloon Facility and the National Science Foundation for their 
excellent support of the Antarctic campaign.


\begin{thebibliography}{99}
\bibitem{Livetime} This relatively low livetime was caused by 
a software bug leading to a slow readout of the GPS time for each event.
\bibitem{LowBandNoise} The band from 230-350~MHz had 
higher $T_{sys}$ and is excluded here due to the low precision of the measurement.
\end{thebibliography}
\end{document}